\documentclass{emulateapj}
\usepackage{apjfonts}
\slugcomment{Accepted by \textsc{the Astrophysical Journal}}

\newcommand{\ddt}[1]{\frac{d #1}{d t}}
\newcommand{\zh}{Z_{\mathrm{H}}}
\newcommand{\sh}{\Sigma_{\mathrm{H}}}
\newcommand{\she}{\Sigma_{\mathrm{He}}}
\newcommand{\estarh}{E^{*}_{\mathrm{H}}}
\newcommand{\estarhe}{E^{*}_{\mathrm{He}}}

\newcommand{\epsh}{\epsilon_{\mathrm{H}}}
\newcommand{\epshe}{\epsilon_{\mathrm{He}}}

\newcommand{\xout}{X_{0}}

\newcommand{\zout}{Z_{0}}

\newcommand{\sdot}{\dot{\Sigma}}

\newcommand{\thy}{T_{\mathrm{H}}}
\newcommand{\thel}{T_{\mathrm{He}}}
\newcommand{\fhy}{F_{\mathrm{H}}}
\newcommand{\fhe}{F_{\mathrm{He}}}
\newcommand{\fb}{F_{\mathrm{base}}}

\newcommand{\cp}{C_{p}}

\newcommand{\themthf}{\thel^{4}-\thy^{4}}

\newcommand{\eq}[1]{#1^{\mathrm{eq}}}
\newcommand{\delsig}{\Delta \Sigma}
\newcommand{\lacc}{l_{\mathrm{acc}}}
\newcommand{\llacc}{\log(\lacc)}

\begin{document}

\title{Hydrogen-Triggered Type I X-ray Bursts in a Two-Zone Model}
\shorttitle{Hydrogen-Triggered X-ray Bursts}

\author{Randall L.\ Cooper and Ramesh Narayan}
\shortauthors{Cooper \& Narayan}
\affil{Harvard-Smithsonian Center for Astrophysics, 60 Garden 
Street, Cambridge, MA 02138}

\email{rcooper@cfa.harvard.edu, rnarayan@cfa.harvard.edu}

\begin{abstract}

We use the two-zone model of Cooper \& Narayan to study the onset and
time evolution of hydrogen-triggered type I X-ray bursts on accreting
neutron stars.  At the lowest accretion rates, thermally unstable
hydrogen burning ignites helium as well and produces a mixed hydrogen
and helium burst.  For somewhat higher accretion rates, thermally
unstable hydrogen burning does not ignite helium and thus triggers
only a weak hydrogen flash.  For our choice of model parameters, these
weak hydrogen flashes occur for $10^{-3} \lesssim \dot{M} /
\dot{M}_{\mathrm{Edd}} \lesssim 3 \times 10^{-3}$.  The peak
luminosities of weak hydrogen flashes are typically much lower than
the accretion luminosity.  These results are in accord with previous
theoretical work.  We find that a series of weak hydrogen flashes
generates a massive layer of helium that eventually ignites in an
energetic pure helium flash.  Although previously conjectured, this is
the first time such bursting behavior has been actually demonstrated
in a theoretical model.  For yet higher accretion rates, hydrogen
burning is thermally stable and thus steadily generates a layer of
helium that ultimately ignites in a pure helium flash.  We find that,
for a narrow range of accretion rates between the mixed hydrogen and
helium burst and weak hydrogen flash regimes, unstable hydrogen
burning ignites helium only after a short series of weak hydrogen
flashes has generated a sufficiently deep layer of helium.  These
bursts have fluences that are intermediate between those of normal
mixed hydrogen and helium bursts and energetic pure helium flashes.

\end{abstract} 


\section{Introduction}

Type I X-ray bursts are thermonuclear explosions that occur on the
surfaces of accreting neutron stars in low-mass X-ray binaries
\citep{Betal75,GH75,Getal76,BCE76,WT76,J77,MC77,LL77,LL78}.  They are
triggered by unstable hydrogen or helium burning \citep[for reviews,
see][]{LvPT93,LvPT95,C04,SB06}.  Theoretical studies of the onset of
type I X-ray bursts suggest that there are in general three different
bursting regimes separated by accretion rate \citep{FHM81,FL87,NH03}.
At the highest accretion rates, helium ignites in a hydrogen-rich
layer and triggers a mixed hydrogen and helium burst.  At somewhat
lower accretion rates, nuclear burning via the hot CNO cycle depletes
hydrogen before helium ignites, and so thermally unstable helium
burning triggers a pure helium flash.  At the lowest accretion rates,
hydrogen burning itself is thermally unstable \citep{CJ80,ET80}.

The vast majority of the type I X-ray bursts that have been observed
so far are mixed hydrogen and helium bursts from systems accreting at
relatively high rates \citep{vPPL88,Cetal03,RLCN06,Getal06}.  This is
perhaps to be expected, for the predicted type I X-ray burst rate is
in general an increasing function of the accretion rate, and
theoretical models predict that the range of accretion rates in which
mixed hydrogen and helium bursts occur is rather large.  Recently,
\citet{GC06} observed pure helium bursts from the accretion-powered
millisecond pulsar SAX J1808.4$-$3658 \citep{intZetal98b}.  These are
the first observed type I X-ray bursts to be unambiguously associated
with the pure helium flash regime.  The Wide Field Cameras onboard the
{\it BeppoSAX} satellite observed nine systems that exhibited type I
X-ray bursts, but from which no persistent emission was detected
\citep{intZetal98,Cetal99,Cetal01,Ketal00,Cetal02}.  The lack of
persistent emission suggests that these ``burst-only'' sources have
accretion rates $\dot{M} \lesssim 0.01 \dot{M}_{\mathrm{Edd}}$, where
$\dot{M}_{\mathrm{Edd}}$ denotes the mass accretion rate at which the
accretion luminosity is equal to the Eddington limit \citep[for a
review, see][]{Cetal04b}.  The low accretion rates imply that the type
I X-ray bursts from these sources likely occur in the thermally unstable
hydrogen burning regime.

Although it is well understood that hydrogen burning is thermally
unstable on neutron stars that accrete at low rates, it is not
immediately apparent what will happen after hydrogen ignites.  While
thermally unstable hydrogen burning will usually ignite helium as well
and produce a mixed hydrogen and helium burst, previous investigations
have found that there exists a range of accretion rates in which
thermally unstable hydrogen burning does not trigger unstable helium
burning and therefore produces only a weak hydrogen flash
\citep{ET80,FHM81,FL87,PBT07}.  The low peak X-ray luminosity of the
weak hydrogen flashes could possibly render them undetectable.  \citet{FL87}
and \citet{PBT07} hypothesize that, within this range of accretion
rates, a series of weak hydrogen flashes will produce a sizable layer
of nearly pure helium that will eventually trigger an energetic helium
flash.

In this investigation, we use a suitably modified version of the
two-zone burst model of \citet{CN06} to study hydrogen-triggered type
I X-ray bursts at low accretion rates.  We find that, for a certain
range of accretion rates, thermally unstable hydrogen burning does not
ignite helium, in agreement with previous investigations.  More
importantly, we find that, within the aforementioned range of
accretion rates, a series of weak hydrogen flashes indeed generates a
layer of helium that eventually ignites in an energetic helium flash,
thereby confirming the hypothesis of \citet{FL87} and \citet{PBT07}.
We begin with a description of the modifications to the two-zone model
in \S \ref{themodel}.  We discuss the equilibria of the governing set
of differential equations and analyze their stability as a function of
accretion rate in \S \ref{eqandstab}.  In \S \ref{integrations} we
integrate the governing equations to determine the nature of type I
X-ray bursts at different accretion rates.  We then summarize our
results and conclude in \S \ref{conclusions}.

\section{The Model}\label{themodel}

We assume that matter accretes spherically onto a neutron star of
gravitational mass $M = 1.4 M_{\odot}$ and areal radius $R = 10 \,
\mathrm{km}$ at an accretion rate per unit area $\sdot$ as measured in
the local frame of the accreted plasma.  We consider all physical
quantities to be functions of the column depth $\Sigma$, which we
define as the rest mass of the accreted matter as measured from the
stellar surface divided by $4 \pi R^{2}$.  We describe the composition
of the matter by the hydrogen mass fraction $X$, helium mass fraction
$Y$, and heavy element fraction $Z = 1-X-Y$.  The mass fractions at
$\Sigma=0$, $X_{0}$, $Y_{0}$, and $Z_{0}$, are those of the accreted
plasma.  In this work, we assume that all heavy elements are CNO and
that the composition of the accreted matter is that of the Sun: $\xout
= 0.7$, $Y_{0} = 0.28$, and $\zout = 0.02$.

To investigate the stability and time evolution of the accreted
plasma, we use the two-zone model of \citet{CN06} which consists 
of (1) a zone that begins
at the surface of the star, where $\Sigma=0$, and extends to the depth
$\sh$ at which hydrogen is depleted via nuclear burning, and (2) a
zone that begins at $\sh$ and extends to the depth $\she$ at which
helium is depleted via nuclear burning.  Hydrogen, helium, and CNO are
all present in zone 1, while only helium and CNO are present in zone
2.  Furthermore, we include both hydrogen and helium burning in zone
1, but we include only helium burning in zone 2.  $\thy$ and $\zh$
denote the temperature and CNO mass fraction at $\sh$, respectively,
and $\thel$ denotes the temperature at $\she$.  For the 
present study we introduce three
modifications to the original two-zone model.  (1) For the low
accretion rates considered in this work, the temperatures in the
accreted layer are often below $8 \times 10^{7}$ K.  At these low
temperatures, $^{13}$N $\beta$-decay competes favorably with
$^{13}$N($p$,$\gamma$)$^{14}$O, and so hydrogen burns predominantly
via the cold CNO cycle \citep{WGS99}
$^{12}$C($p$,$\gamma$)$^{13}$N($\beta^{+}\nu$)$^{13}$C($p$,$\gamma$)$^{14}$N($p$,$\gamma$)$^{15}$O($\beta^{+}
\nu$)$^{15}$N($p$,$\alpha$)$^{12}$C, the rate of which is determined
primarily by the slow temperature-dependent reaction
$^{14}$N($p$,$\gamma$)$^{15}$O \citep{APJ92}.  Therefore, we use the
more general prescription for the hydrogen nuclear energy generation
rate $\epsh$ described in \citet{NH03}, which includes cold CNO cycle
hydrogen burning, with the exceptions that the
$^{13}$N($p$,$\gamma$)$^{14}$O and $^{14}$N($p$,$\gamma$)$^{15}$O
reaction rates are updated to those of \citet{CF88}.  (2) In this
work, the outward heat flux due to non-equilibrium electron captures,
neutron emissions, and pycnonuclear reactions that occur in the both
the outer and inner crust \citep{HZ90v227,HZ03,B00,GBSMK06} largely
sets the thermal structure of the accreted layer \citep{FHIR84,HF84}.
This is in contrast to the accreted layers on neutron stars that
accrete at high rates, for which mixed hydrogen and helium
burning sets the thermal structure.  Therefore, we follow \citet{B00}
and set the outward flux at the base of the accreted layer due to deep
crustal heating $\fb = 0.1 \,\mathrm{MeV} (\sdot/m_{\mathrm{u}})$,
where $m_{\mathrm{u}}$ is the atomic mass unit.  (3) In their
derivation of the two differential equations that govern the time
evolution of the temperature $\thel$ and column depth $\she$ of zone
2, the helium-burning zone, \citet{CN06} assumed that instantaneous
changes of the physical quantities $\sh$, $\zh$, and $\thy$ of zone 1
affect the physical quantities $\she$ and $\thel$ of zone 2.  This
assumption was physically well-motivated for their work because the
thermal and accretion timescales of zone 1 were on the order of the
thermal and accretion timescales of zone 2, respectively.  As we show
in \S \ref{eqandstab}, however, the two timescales of zone 2 in this
work are much greater than their counterparts in zone 1, so the
previously stated assumption is no longer valid.  Therefore, we drop
this assumption when we derive the differential equations that govern
the current model.  The result of the new derivation is equivalent to
simply omitting the $d\sh/dt$, $d\zh/dt$, and $d\thy/dt$ terms that
appear in equations (32) and (33) of \citet{CN06}.  The fundamental
equations of our two-zone model are thus
\begin{equation}\label{dshdteqn}
\ddt{\sh} = 2 \left [\sdot - \frac{\epsh(\zh,\thy)}{\xout \estarh}\sh \right ],
\end{equation}
\begin{equation}\label{dzhdteqn}
\ddt{\zh} = 2 \frac{\epshe(\thy)}{\estarhe} -
\frac{(\zh-\zout)}{\sh}\left (2 \sdot-\ddt{\sh} \right ),
\end{equation}
\begin{equation}\label{dthdteqn}
\ddt{\thy} = \frac{5}{4 \cp} \left [\epsh(\zh,\thy) +\epshe(\thy) - \frac{\fhy - \fhe}{\sh} \right ] +
\frac{\thy}{4 \sh} \ddt{\sh},
\end{equation}
\begin{equation}\label{dshedteqn}
\ddt{\she} = 2 \left [\sdot -\frac{\epshe(\thel)}{(1-\zh) \estarhe} \delsig \right ],
\end{equation}
\begin{eqnarray}\label{dthedteqn}
\ddt{\thel} = \frac{5 (\themthf)^{2}}{4\cp (\thel^{8} -5 \thel^{4}
\thy^{4} + 4 \thel^{3}\thy^{5})} \left [\epshe(\thel) -
\frac{\fhe-\fb}{\delsig}\right ] \nonumber \\+
\frac{\thel}{4\delsig}\left[1-\left(\frac{\thy}{\thel}\right)^{4}\right]
\ddt{\she} \nonumber \\.
\end{eqnarray}
See \citet{CN06} for the definitions of the various symbols.  

We emphasize that, in an effort to construct a model that is as simple as 
possible and yet contains all of the necessary physics needed to 
describe hydrogen-triggered type I X-ray bursts, we make several 
assumptions that are not always valid.  First, we follow 
\citet{CN06} and assume an ideal gas equation of state, but the electrons 
are degenerate and set the equation of state at large 
column depths.  In the governing equations (\ref{dshdteqn}-\ref{dthedteqn}) 
however, the equation of state of the matter shows up only in the
specific heat at constant pressure $\cp = 5k_{\mathrm{B}}/2m_{p}$.  The ion
specific heat should dominate that of the electrons for the range of
column depths we consider, so our approximation is reasonable.  Second, 
we ignore the density dependence of the helium nuclear energy generation 
rate $\epshe$, and so the rate is somewhat overestimated 
in zone 1 and underestimated in zone 2.  However, considering the other 
approximations we have made it should be accurate enough for our purposes.
Third, we set the opacity to the constant value $\kappa = 0.136$
$\mathrm{cm}^{2}\,\mathrm{g}^{-1}$.  This approximation is probably the 
most severe.  Electron conduction sets the opacity at the largest column 
depths we consider in 
this work, and so the true opacity can be notably smaller than we 
assume.  We discuss the consequences of this approximation in \S 
\ref{integrations}.

\section{Equilibria and Their Stability}\label{eqandstab}

\begin{figure}
\epsscale{1.0}
\plotone{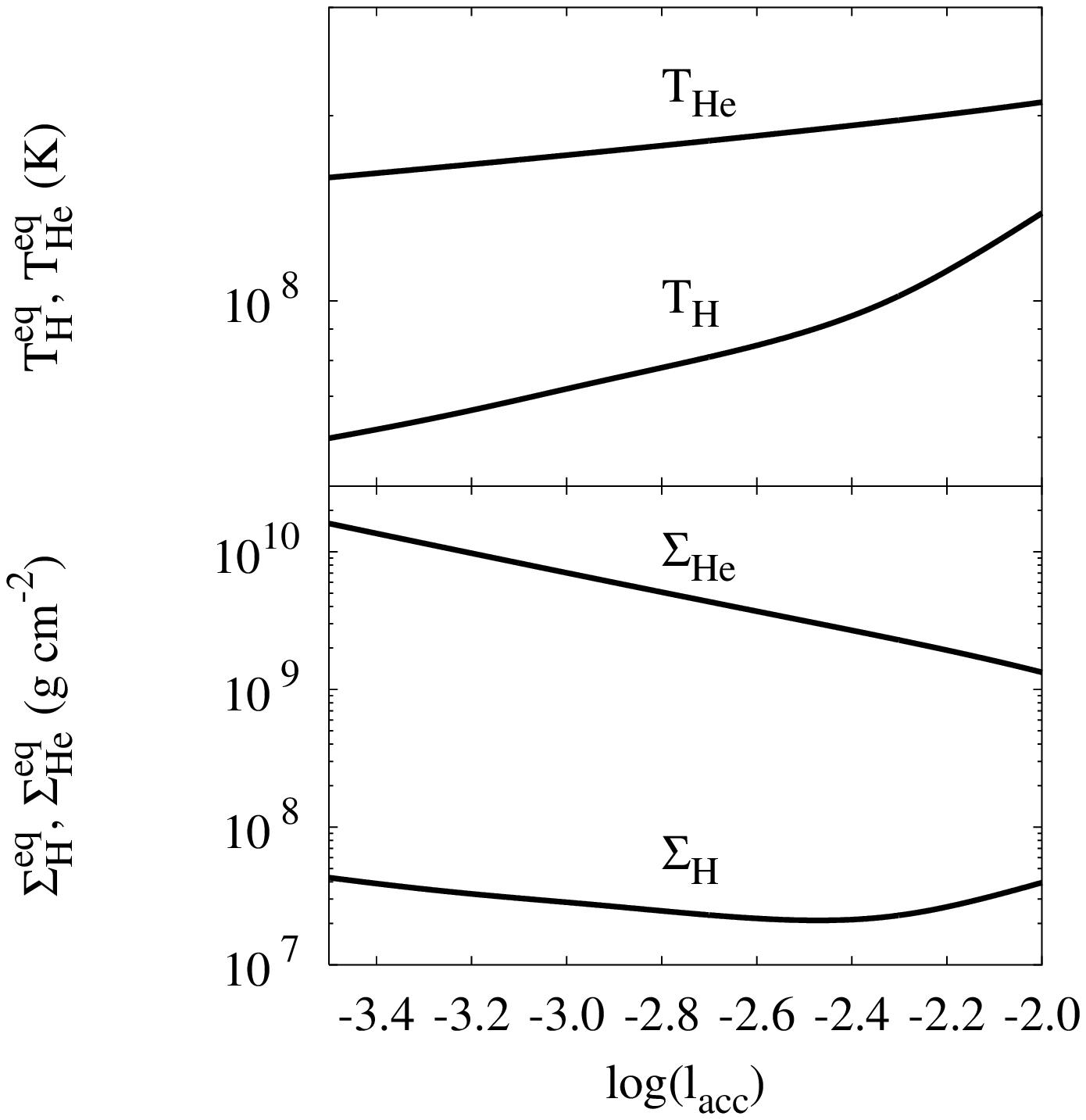}
\caption{Equilibrium values of the four variables $\thy$, $\thel$,
$\sh$, and $\she$ as a function of $\lacc \equiv \sdot /
\sdot_{\mathrm{Edd}}$.  $\eq{\zh} 
\approx \zout$ for the entire range of $\lacc$ considered in this work.  
$\eq{\sh}$ reaches a local minimum at roughly the same $\lacc$ at 
which $\eq{\thy}$ is great enough to initiate hot CNO cycle hydrogen 
burning.}
\label{equilibria}
\end{figure}

In this section, we solve for the equilibrium solutions of the
two-zone model and determine their stability.  To find the
equilibria, we set $d/dt = 0$, so that equations
(\ref{dshdteqn}-\ref{dthedteqn}) become the following set of five coupled 
algebraic equations: 
\begin{equation}\label{eqeqn1}
\epsh(\zh,\thy) \sh = \sdot \xout \estarh,
\end{equation}
\begin{equation}
\epshe(\thy) \sh = \sdot (\zh-\zout)\estarhe,
\end{equation}
\begin{equation}
\fhy-\fhe = [\epsh(\zh,\thy) + \epshe(\thy)]\sh,
\end{equation}
\begin{equation}
\epshe(\thel)  \delsig = \sdot (1-\zh)\estarhe,
\end{equation}
\begin{equation}\label{eqeqn5}
\fhe -\fb = \epshe(\thel)\delsig.
\end{equation}
We solve equations (\ref{eqeqn1}-\ref{eqeqn5}) for the five
equilibrium values $\eq{\sh}$, $\eq{\zh}$, $\eq{\thy}$, $\eq{\she}$,
and $\eq{\thel}$.  When we do this, we obtain
\begin{equation}
\eq{\fhy} = \fb+\sdot \left[ \xout \estarh + (1-\zout)\estarhe \right ],
\end{equation}
\begin{equation}
\eq{\fhe} = \fb+\sdot (1-\eq{\zh})\estarhe.
\end{equation}
Thus the flux emitted from the stellar surface, $\eq{\fhy}$, equals
the flux released via steady-state nuclear burning of all the fuel plus 
the flux $\fb$ due to deep crustal heating,
and the flux entering zone 1, $\eq{\fhe}$, equals the flux released
via steady-state nuclear burning of the helium within zone 2 plus 
that due to deep crustal heating.  
See Figure \ref{equilibria} for plots of $\eq{\thy}$, $\eq{\thel}$, $\eq{\sh}$, and $\eq{\she}$ as a function of the Eddington-scaled
accretion rate
\begin{equation}
\lacc \equiv \frac{\sdot}{\sdot_{\mathrm{Edd}}},
\end{equation}
where $\sdot_{\mathrm{Edd}} = 1.0 \times 10^{5}$
$\mathrm{g}\,\mathrm{cm}^{-2}\,\mathrm{s}^{-1} $.  

The trends seen in Figure \ref{equilibria} may be understood as
follows.  Hydrogen and helium burning generates energy and heats the
accreted layer.  In equilibrium, hydrogen and helium burn at a rate
$\propto \lacc$, and so the equilibrium temperatures $\eq{\thy}$ and
$\eq{\thel}$ are monotonically increasing functions of $\lacc$.  For
$\llacc \lesssim -2.5$, the rates at which both hydrogen and helium
burn are very temperature-sensitive, and so the depths at which
hydrogen and helium deplete, $\eq{\sh}$ and $\eq{\she}$, are
decreasing functions of $\lacc$.  For $\eq{\thy} \gtrsim 8 \times
10^{7}$ K, however, the proton capture rate onto $^{13}$N($t_{1/2} =
598$ s) exceeds that of its $\beta$-decay, and so hydrogen burning
proceeds via the hot CNO cycle
$^{12}$C($p$,$\gamma$)$^{13}$N($p$,$\gamma$)$^{14}$O($\beta^{+}
\nu$)$^{14}$N($p$,$\gamma$)$^{15}$O($\beta^{+}
\nu$)$^{15}$N($p$,$\alpha$)$^{12}$C, the rate of which is determined
by the slow $\beta$-decays of $^{14}$O($t_{1/2}=70.6$ s) and
$^{15}$O($t_{1/2}=122$ s) \citep{HF65}.  In this case, the hydrogen
burning rate depends only on the CNO mass fraction $\zh$, which is
$\approx \zout$ because $\epshe(\eq{\thy})/\estarhe$, the rate at
which helium burns to carbon, is negligible.  Equation (\ref{eqeqn1})
thus implies that $\eq{\sh} \propto \sdot$.  This regime is reflected
in the positive slope of $\eq{\sh}(\lacc)$ in Figure \ref{equilibria}.

\begin{figure}
\plotone{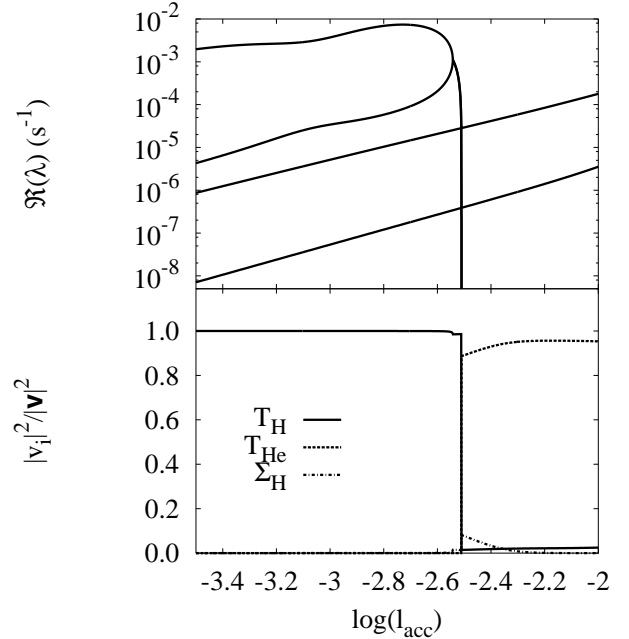}
\caption{Upper panel: real parts of the positive eigenvalues as a
function of $\lacc$.  Lower panel: normalized components of the
eigenvector corresponding to the eigenvalue with the greatest real
part.  The $\zh$ and $\she$ components are negligible and thus are not
shown.  Both the eigenvalues and eigenvectors show that perturbations
in $\thy$ trigger bursts for $\llacc \lesssim -2.5$, and perturbations
in $\thel$ trigger bursts for $\llacc \gtrsim -2.5$.}
\label{evalandvec}
\end{figure}

After we have solved for the equilibrium solution, we conduct a linear
stability analysis by assuming that all quantities vary with 
time as $\exp(\lambda t)$ as described in \citet{CN06}.  The top panel of
Figure \ref{evalandvec} shows the spectrum of the positive real parts
of the eigenvalues $\lambda$, and the bottom panel of Figure
\ref{evalandvec} shows the normalized squared moduli of the components
of the eigenvector ${\bf v}$ corresponding to the eigenvalue with the
greatest real part, both as a function of $\lacc$.  

One can ascertain the physical meanings of the eigenvalue spectrum by
considering perturbations of proper subsets of the system of equations
(\ref{dshdteqn}-\ref{dthedteqn}).  A linear stability analysis of only
equations (\ref{dshdteqn}) and (\ref{dthdteqn}), which describe
perturbations of the hydrogen-burning shell, produces two eigenvalues
that are virtually indistinguishable from the two largest eigenvalues
of the full system.  Similarly, a linear stability analysis of only
equations (\ref{dshedteqn}) and (\ref{dthedteqn}), which describe
perturbations of the helium-burning shell, produces two eigenvalues
that are virtually indistinguishable from the two lower eigenvalues
(which vary nearly linearly with $\lacc$ in Figure \ref{evalandvec}).
Thus the two largest eigenvalues that abruptly become negative at
$\llacc \approx -2.5$ describe the stability of hydrogen burning in
zone 1, and the other two nearly linear eigenvalues describe the
stability of helium burning in zone 2.  The clear distinction between
the two sets of eigenvalues implies that, according to the linear
stability analysis, there is little interaction between zone 1 and
zone 2.  Physically, this is because $\eq{\sh} \ll \eq{\she}$,
i.e. the two burning fronts are well-separated.  One can deduce this
also from the $5 \times 5$ Jacobian matrix of the full linear
stability analysis.  The Jacobian is nearly a block diagonal matrix,
with a $3 \times 3$ block matrix in the upper left corner that
corresponds to the Jacobian of equations
(\ref{dshdteqn}-\ref{dthdteqn}) and a $2 \times 2$ block matrix in the
lower right corner that corresponds to the Jacobian of equations
(\ref{dshedteqn}-\ref{dthedteqn}).  Although the two off-diagonal
block matrices are nonzero in general, their matrix elements are
negligible relative to those of the two block diagonal matrix
elements, which suggests that the full system is reducible to 
a set of two essentially independent systems.


Only the mode corresponding to
the eigenvalue with the greatest real part determines the behavior of
the system near the equilibrium \citep[e.g.,][]{GH83}.  From a quick
examination of the largest eigenvalue and its corresponding
eigenvector shown in Figure \ref{evalandvec}, our linear stability
analysis suggests that thermally unstable hydrogen burning in zone 1
triggers bursts for $\llacc \lesssim -2.5$ and thermally unstable
helium burning in zone 2 triggers bursts for $\llacc \gtrsim -2.5$.
This agrees well with the results of previous theoretical
investigations \citep{FHM81,FL87,NH03}.  Note that the value of
$\lacc$ at which this transition occurs is in general a decreasing
function of $\fb /\sdot$, which is assumed to be $(0.1 \,\mathrm{MeV})
/m_{\mathrm{u}}$ in this work.

Figure \ref{equilibria} shows that $\eq{\thy} \gtrsim 8 \times
10^{7}$ K for $\llacc \gtrsim -2.5$, and so hydrogen burns via 
the temperature-independent hot CNO cycle.  The hydrogen-burning 
layer is thus stable to thermal perturbations, and so the 
eigenvalues corresponding to the hydrogen-burning layer become 
negative.  The thermal eigenvalue of the hydrogen-burning layer 
passes through zero at approximately the same $\lacc$ at which 
$\eq{\sh}$ reaches a local minimum \citep[e.g.,][]{P83}.

\section{Numerical Integrations of the Two-Zone Model}\label{integrations}

The linear stability analyses presented in \S \ref{eqandstab} describe
the behavior of the dynamical systems near their respective
equilibria.  However, the time evolution of a system that is
dynamically unstable need not come anywhere close to its equilibrium.
This is particularly important for systems in which hydrogen burning
is thermally unstable.  In this regime, a thermonuclear instability
ensues when the column depth of the accreted layer $\Sigma \approx
\eq{\sh} \ll \eq{\she}$, and so the behavior of $\she$ cannot
necessarily be deduced from the stability analysis of small 
perturbations around the equilibrium.  Moreover, the
linear stability analysis suggests that, in this regime, helium
burning in zone 2 is unstable as well, which makes the true
time-dependent behavior of the systems especially difficult to predict
from the stability analysis alone.  In this section, we remedy this by
numerically integrating equations (\ref{dshdteqn}-\ref{dthedteqn}) to
study the nonlinear development of the instability and the limit cycle
behavior of type I X-ray bursts at different accretion rates.

\begin{figure}
\plotone{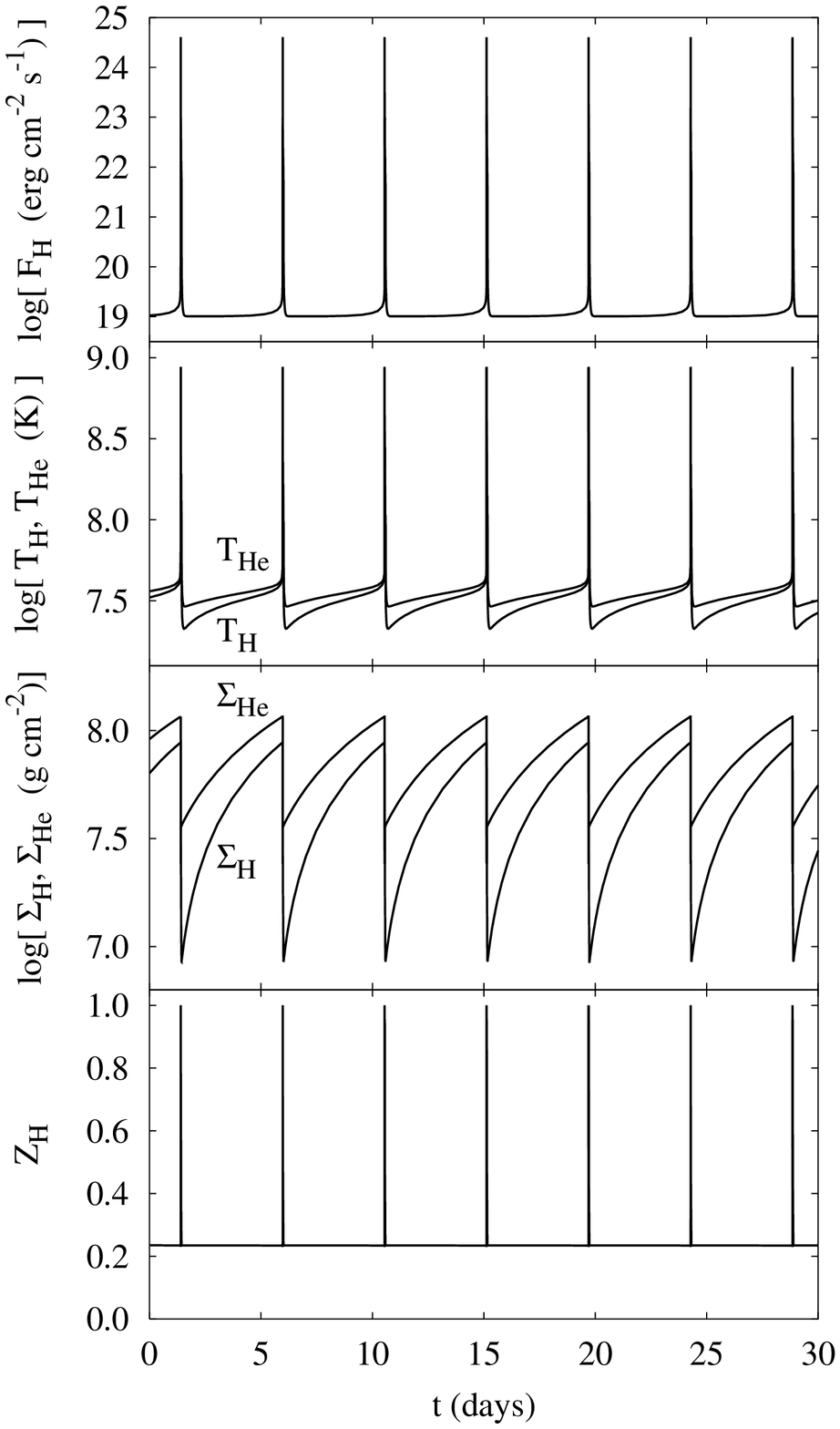}
\caption{Time evolution of type I X-ray bursts for $\llacc = -3.0$.
The top panel shows the light curve, and the bottom three panels show
the time evolution of the five physical quantities.  Note that $\thel
> \thy$ and $\she > \sh$.  Sudden drops in both $\sh$ and $\she$
during bursts indicate that both hydrogen and helium have ignited and
thus have triggered a mixed hydrogen and helium burst.}
\label{Hint3.0}
\end{figure}

We initiate all numerical integrations by perturbing the equilibrium
solution, and we continue the integration until the system reaches a
limit cycle.  The limit cycle behavior is completely insensitive to
the initial conditions.  Figure \ref{Hint3.0} shows the limit cycle
behavior for a system with $\llacc = -3.0$.  At the lowest accretion
rates we consider in this work, unstable hydrogen burning via the cold
CNO cycle raises the temperature of the accreted layer to such an
extent as to trigger unstable helium burning as well.  The sudden
drops in both $\sh$ and $\she$ that correspond with sudden increases
in the temperatures $\thy$ and $\thel$ and the outward flux $\fhy$
indicate that both hydrogen and helium ignite in every burst.  These
bursts are the well-known mixed hydrogen and helium type I X-ray
bursts triggered by thermally unstable hydrogen burning
\citep{ET80,FHM81,FL87,PBT07}.

\begin{figure*}
\epsscale{1.0} 
\plottwo{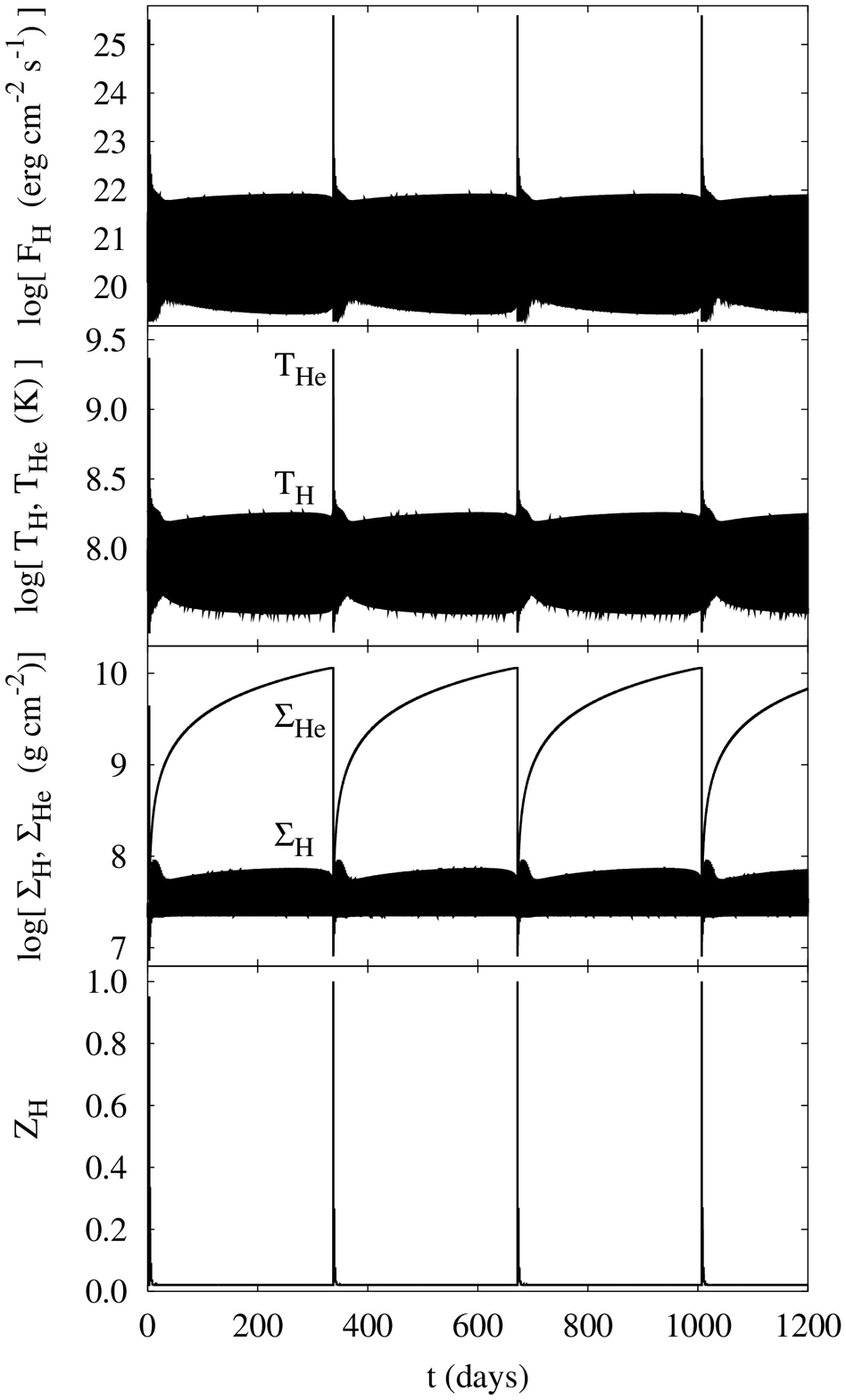}{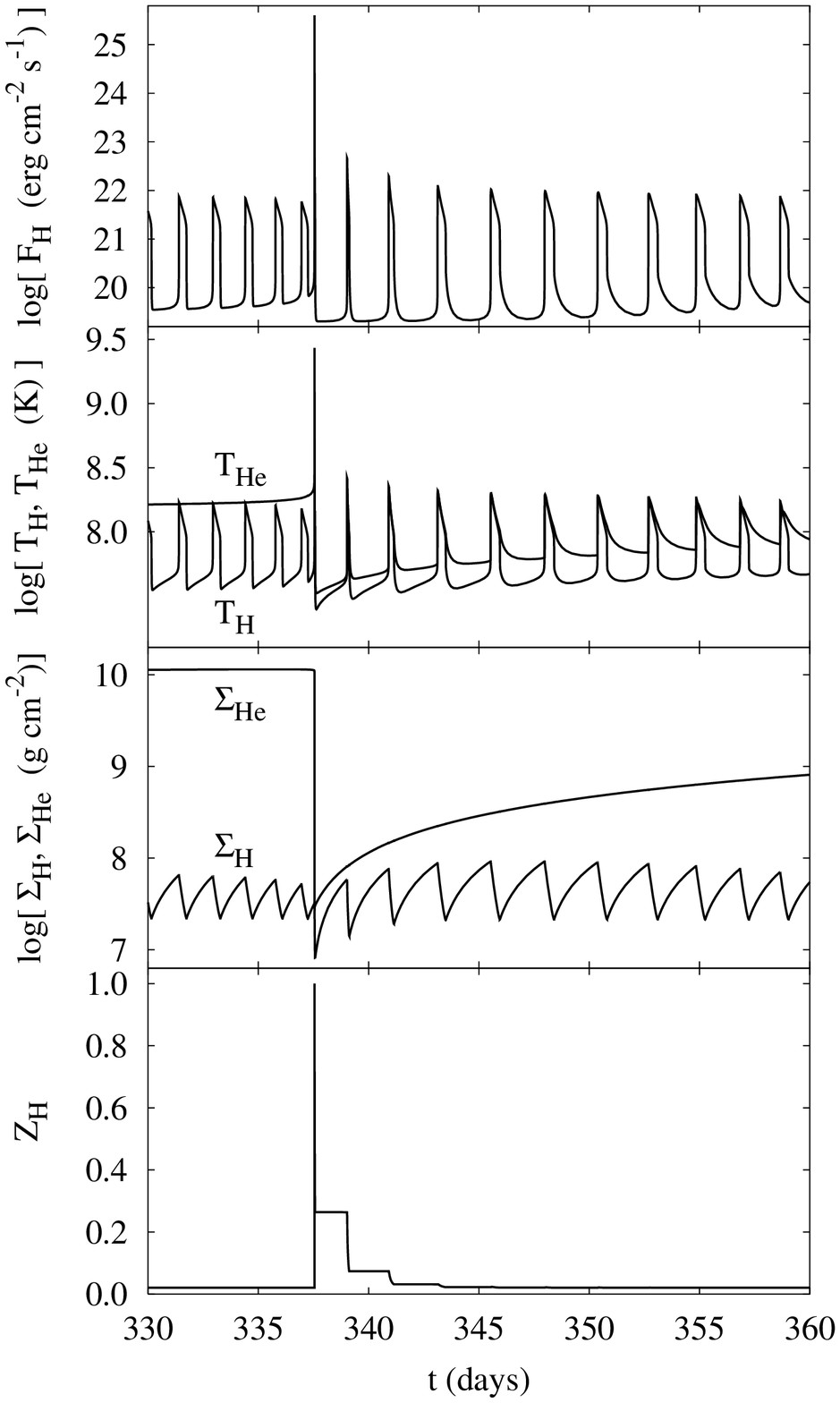}
\caption{Time evolution of bursts for $\llacc = -2.7$.  The top panel
on the left shows the light curve, and the bottom three panels show
the time evolution of the five physical quantities.  Note that $\thel
> \thy$ and $\she > \sh$.  The increase in $\thy$ due to thermally
unstable hydrogen burning is insufficient to trigger helium ignition.
Therefore, $\she$ increases during a long series of weak hydrogen
flashes until helium ignites and produces an energetic pure helium
flash.  The set of panels on the right illustrates the time evolution
of the system just before and after an energetic pure helium flash.}
\label{Hint2.7}
\end{figure*}

For somewhat higher accretion rates, theory predicts that the increase
in temperature due to thermally unstable hydrogen burning is
insufficient to ignite helium.  Previous authors have speculated that
a series of weak hydrogen flashes allows for the accumulation of a
large layer of helium and eventually leads to a strong helium flash
\citep{T81,FL87,PBT07}.  We present in Figure \ref{Hint2.7} the
results of a calculation for a system with $\llacc = -2.7$ that
confirms this suggestion: if thermally unstable hydrogen burning does
not ignite helium, helium will accumulate up to a column depth $\she
\sim 10^{10}\,\mathrm{g}$ $\mathrm{cm}^{-2}$ and eventually ignite in
an energetic helium flash.  Subsequent weak hydrogen flashes ensue
until a sufficient amount of helium can accumulate and once again
trigger an energetic helium flash.  To our knowledge, this is the
first study in which this limit cycle behavior has been explicitly 
demonstrated.

\begin{figure*}
\epsscale{1.0} 
\plottwo{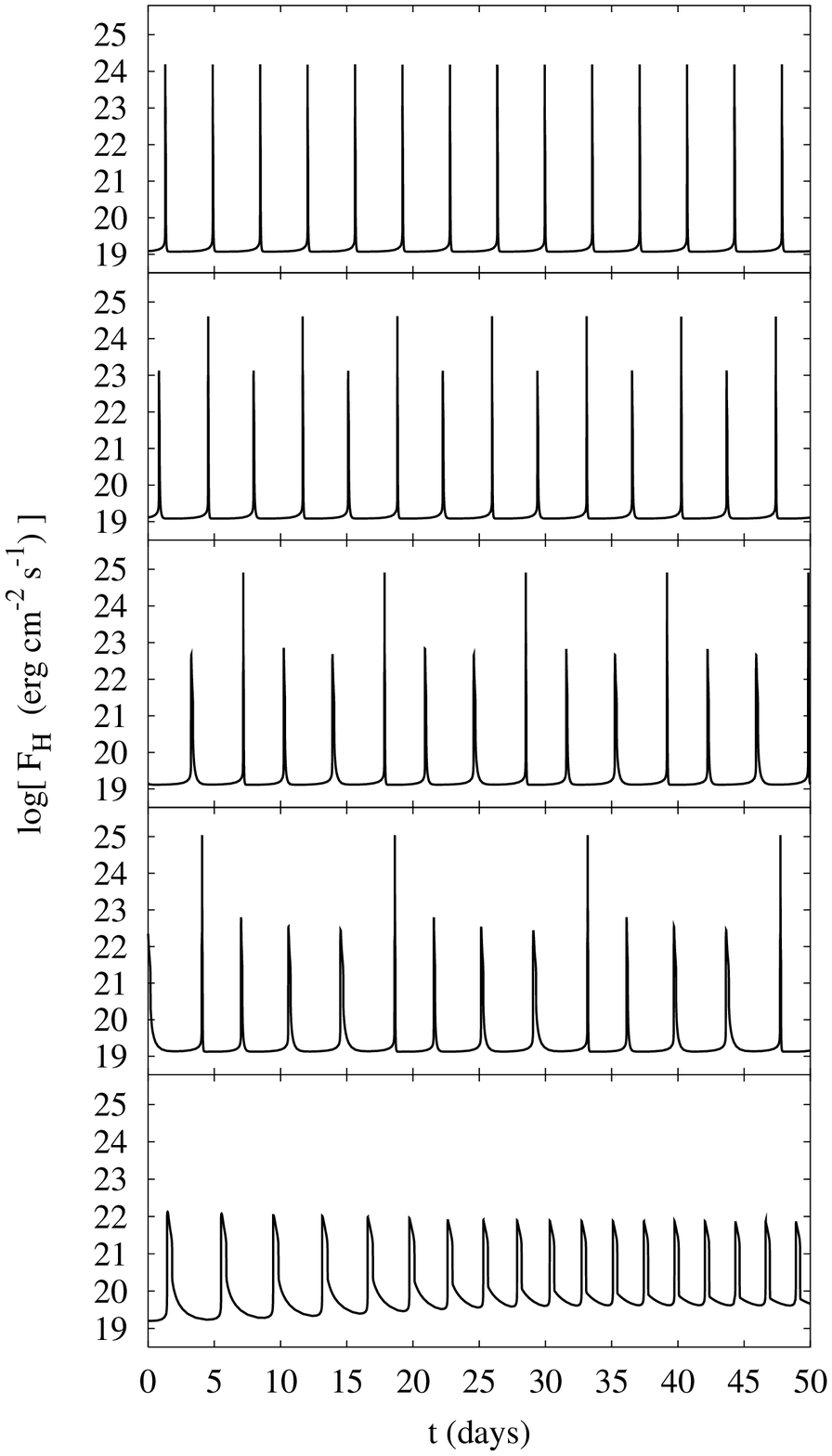}{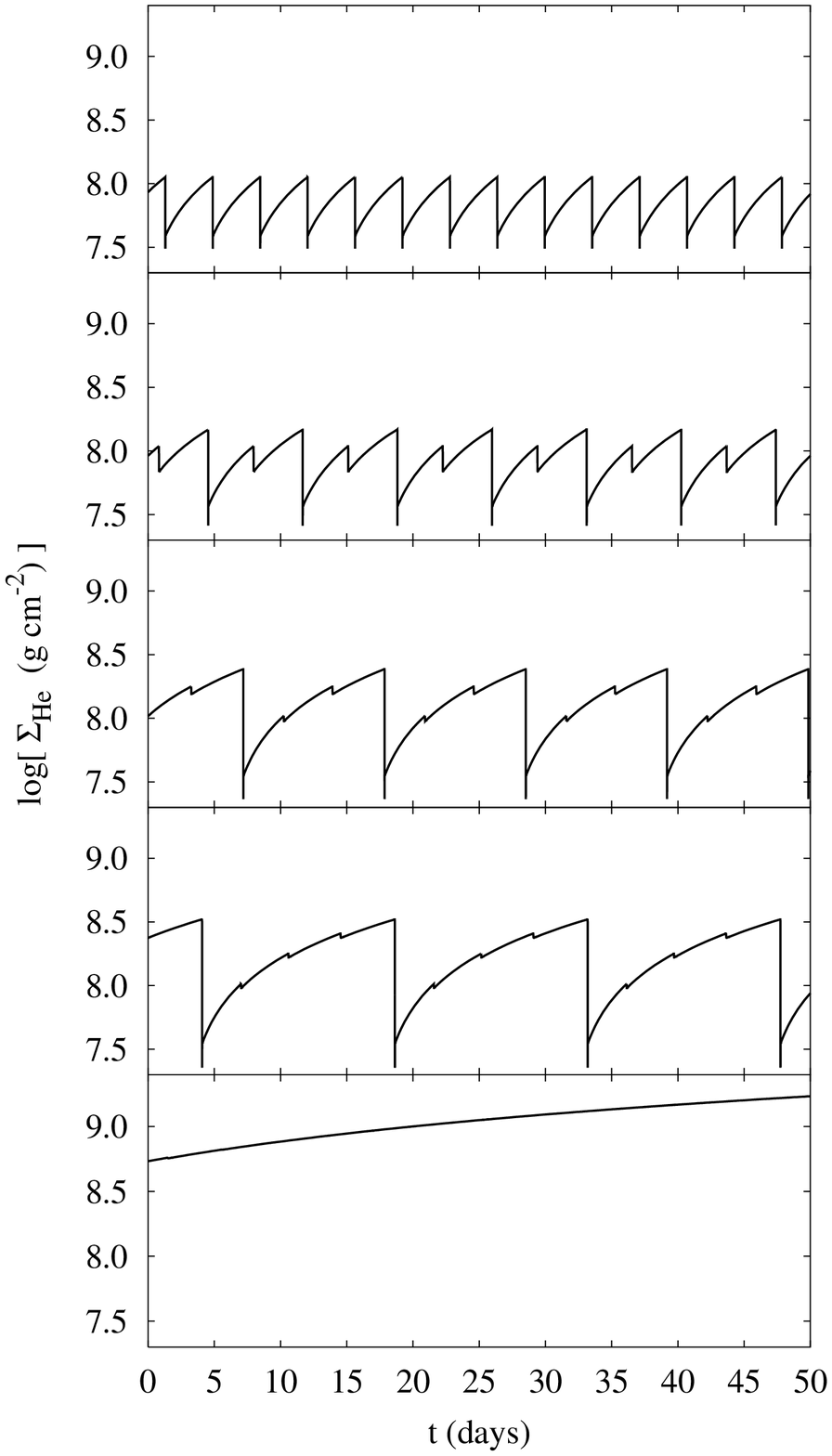}
\caption{Burst light curves (left) and $\she$ (right) for $\llacc =
-2.93$, $-2.92$, $-2.89$, $-2.88$, and $-2.87$, from top to bottom.
For $\llacc = -2.93$, the $\thy$ increase due to thermally unstable
hydrogen burning is always sufficient to ignite all of the helium.
For slightly larger values of $\lacc$, unstable hydrogen burning
triggers complete helium ignition only if $\she$ is sufficiently
large, and the ensuing mixed hydrogen and helium bursts are therefore
less frequent and more energetic.  For $\llacc = -2.87$, unstable
hydrogen burning never ignites helium.}
\label{Hintmulti}
\end{figure*}

The bursting behavior of systems with accretion rates that lie in a
small range between those of the two regimes described above can be
somewhat chaotic. Figure \ref{Hintmulti} shows the light curve $\fhy$ 
and the time evolution of the helium column depth $\she$ for
five different accretion rates that span the transition region between
the two aforementioned bursting regimes.  As we increase the accretion
rate from $\llacc = -2.93$ to $-2.87$, a comparison between $\fhy$
and $\she$ illustrates that the bursting behavior evolves from
that of exclusively mixed hydrogen and helium bursts, to a combination
of both mixed bursts and weak hydrogen flashes, to predominantly weak
hydrogen flashes with infrequent but unusually energetic mixed bursts.
The left panel of Figure \ref{Hintmulti} shows that the amount of
helium fuel burned in the mixed bursts increases with accretion rate.
Thus the burst fluences of the mixed bursts in this transition region
are intermediate between the mixed bursts at slightly lower $\lacc$
and the energetic pure helium bursts and slightly higher $\lacc$.

\begin{figure}
\plotone{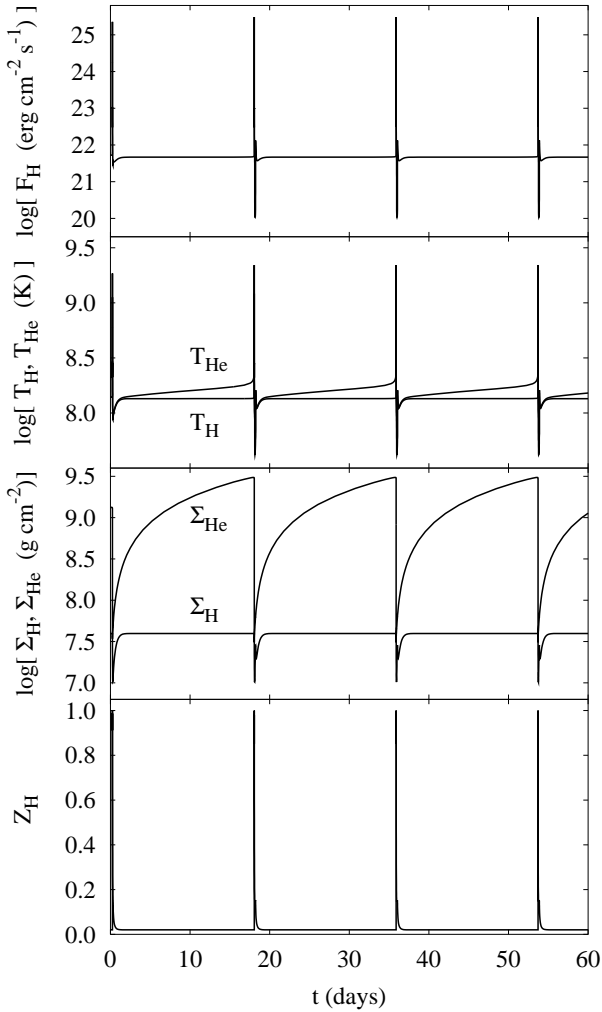}
\caption{Time evolution of bursts for $\llacc = -2.0$.  The top panel
shows the light curve, and the bottom three panels show the time
evolution of the five physical quantities.  $\thy \gtrsim 10^{8}$ K,
so hydrogen burns via the temperature-independent and thus thermally
stable hot CNO cycle.  Consequently, stable hydrogen burning generates
a large and nearly pure helium layer that eventually ignites in an
energetic helium flash.}
\label{Hint2.0}
\end{figure}

For $\llacc \gtrsim -2.5$, Figure \ref{evalandvec} implies that
hydrogen burning is thermally stable.  Consequently, steady-state
hydrogen burning generates a sizable column of nearly pure helium as
accretion ensues until the helium at the base of the accreted layer
ignites.  Figure \ref{Hint2.0} depicts the limit cycle behavior of
such a system with $\llacc = -2.0$.  The existence and nature of these
pure helium-shell flashes are well-established in theoretical models
\citep{FHM81,HF82,FL87,B98,CB00,NH03,Wetal04,CMintZP06}.  Note that
the column depths $\she$ at which helium ignites shown in Figures
\ref{Hint2.7} and \ref{Hint2.0} are generally lower than those found
by \citet{CMintZP06} and \citet{PBT07}.  This is due to our
simplification that the opacity $\kappa = 0.136$
$\mathrm{cm}^{2}\,\mathrm{g}^{-1}$ throughout the entire layer.  At
these large column depths, electron conduction sets the opacity, and
so the true $\kappa$ is lower than our assumed value.  Since the
opacity is lower, a larger column of accreted matter must be accreted
before helium burning can become thermally unstable.

\begin{figure}
\plotone{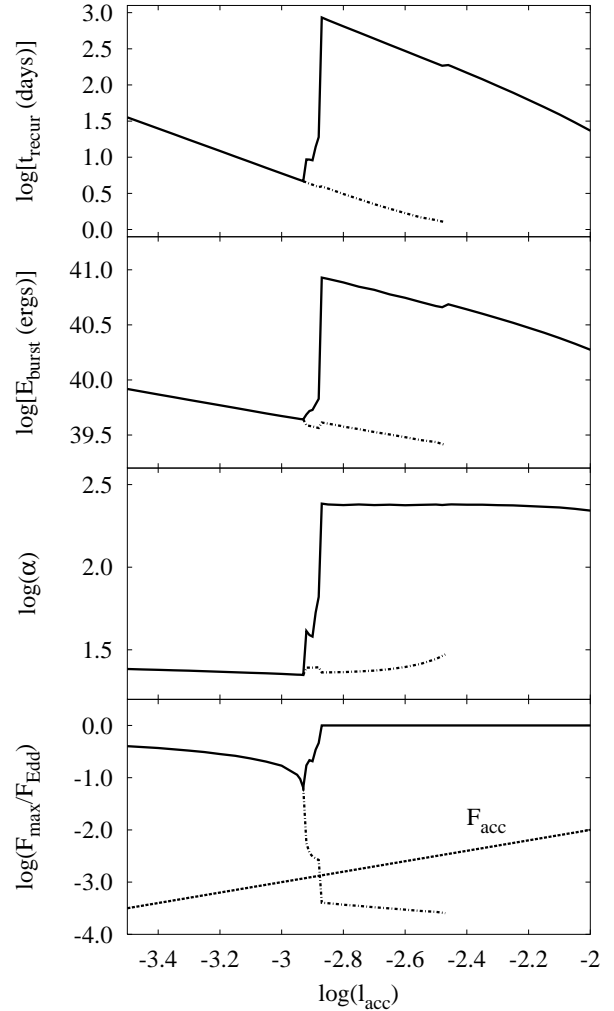}
\caption{Recurrence time, burst energy, ratio of accretion to burst
energy $\alpha$, and peak flux of type I X-ray bursts as a function of
$\lacc$, from top to bottom.  For $-2.9 \lesssim \llacc \lesssim
-2.5$, both energetic pure helium flashes and weak hydrogen flashes
occur.  In this case, the solid line denotes the burst properties of
the helium flashes and the dot-and-dashed line denotes the burst
properties of the hydrogen flashes.  The chaotic behavior discussed in
\S \ref{integrations} and Fig. \ref{Hintmulti} causes the sudden jumps
in the panels that occur in the range $-2.93 < \llacc < -2.87$.  The
dashed line in the bottom panel denotes the luminosity due to
accretion.  Note that the peak flux during most weak hydrogen flashes
is much less than the accretion luminosity.
}
\label{Hburstprops}
\end{figure}

We plot the recurrence time, burst energy, $\alpha$-value, and maximum
Eddington-scaled flux of type I X-ray bursts as a function of $\lacc$
in Figure \ref{Hburstprops}.  All quantities are as measured by an
observer at infinity.  For $-2.9 \lesssim \llacc \lesssim -2.5$, both
energetic pure helium flashes and weak hydrogen flashes occur.  In
this regime, the solid lines in the four panels denote the burst
properties of the helium flashes, and the dot-and-dashed lines denote
the burst properties of the hydrogen flashes.  The recurrence time is
defined as the time between two successive flashes.  We calculate the
burst energy by determining the total amount of hydrogen and helium
fuel that exists just prior to a burst and assuming that all of the
fuel burns during a burst.  Since essentially no helium burns during a
weak hydrogen flash, however, we determine the burst energy of such a
flash by instead assuming that the burst consumes only hydrogen.  The
parameter $\alpha$ is defined as the accretion energy released between
successive bursts divided by the nuclear energy released during a
burst.  For the range of accretion rates in which both energetic pure
helium flashes and weak hydrogen flashes occur, the larger $\alpha$ in
Figure \ref{Hburstprops} is the value an observer would measure if
only the energetic helium flashes were detected, and the smaller
$\alpha$ is the value an observer would measure if the weak hydrogen
flashes were detected as well.  Since the peak flux of weak hydrogen
flashes is usually much less than the accretion luminosity, as is
evident in the bottom panel of Figure \ref{Hburstprops}, it is
possible that weak hydrogen flashes would be undetected.  Note that,
as Figure \ref{Hint2.7} illustrates, the properties of the weak
hydrogen flashes for a given $\lacc$ vary somewhat with time.
Consequently, the quantities plotted in Figure \ref{Hburstprops} that
describe weak hydrogen bursts are only approximate.

The ranges of accretion rates within which the various bursting
regimes lie depend upon the value of the outward flux at the base of
the accreted layer $\fb$, which we set to $0.1 \,\mathrm{MeV}
(\sdot/m_{\mathrm{u}})$.  However, for the low accretion rates we
consider in this work, $\fb$ may be significantly larger \citep{B04}.
When we instead set $\fb = 1.0 \,\mathrm{MeV} (\sdot/m_{\mathrm{u}})$,
we find that weak hydrogen flashes occur for $-3.95 \lesssim \llacc
\lesssim -2.55$.  This range is lower that that shown in Figure
\ref{Hburstprops}, and it is qualitatively consistent with the results
of \citet{PBT07}.

Our numerical integrations expose a fundamental weakness of linear
stability analyses.  While one can accurately determine whether or not
a system is stable to perturbations using a linear stability analysis,
one cannot necessarily infer with any confidence the nature of a
system that is unstable.  For example, numerical integrations
demonstrate that a transition between hydrogen-triggered mixed bursts
and weak hydrogen flashes occurs at $\llacc \approx -2.9$, but neither
the eigenvalues nor the eigenvector shown in Figure \ref{evalandvec}
indicate any change in the physical behavior of the system near this
accretion rate.  This weakness exists even in the more complete global
linear stability analysis of \citet{NH03}.  In that work, the authors
found that hydrogen burning is thermally unstable at low accretion
rates as well, but again they were unable to deduce the nature of the
instability from their results.  Thus, one must use caution when
inferring the time-dependent behavior of a system that is unstable
according to a linear stability analysis.

\section{Summary and Conclusions}\label{conclusions}

Using a suitably modified version of the two-zone model of
\citet{CN06}, we have studied both the onset and time evolution of
hydrogen-triggered type I X-ray bursts on accreting neutron stars.  At
the lowest accretion rates we have considered, thermally unstable
hydrogen burning triggers thermally unstable helium burning and thus
produces a mixed hydrogen and helium burst.  For somewhat higher
accretion rates, thermally unstable hydrogen burning does not ignite
helium and thus triggers only a weak hydrogen flash, in agreement with
previous studies.  For our choice of model parameters, these weak
hydrogen flashes occur for $10^{-3} \lesssim \dot{M} /
\dot{M}_{\mathrm{Edd}} \lesssim 3 \times 10^{-3}$.  We find that, in
accord with the predictions of \citet{FL87} and \citet{PBT07},
subsequent weak hydrogen flashes generate a sizable layer of nearly
pure helium that eventually ignites in an energetic pure helium flash.
This is the first time this bursting behavior has been seen in
theoretical models.  For yet higher accretion rates, hydrogen burning
is thermally stable and thus steadily generates a layer of helium that
ultimately ignites in a pure helium flash.  In addition, we find that
there exists a small range of accretion rates near the boundary
between the mixed hydrogen and helium burst and weak hydrogen flash
regimes where thermally unstable hydrogen burning ignites helium only
after $\she$ is sufficiently large.  The resulting mixed hydrogen and
helium type I X-ray bursts have fluences that are intermediate between
the normal mixed hydrogen and helium bursts and the energetic pure
helium flashes.

The simplicity of our two-zone model has enabled us to survey a wide
range of accretion rates and to illustrate the basic physics of
hydrogen-triggered type I X-ray bursts in the various bursting
regimes.  However, there are several pieces of physics we have not
considered that certainly would affect the nature of type I X-ray
bursts at the low accretion rates considered in this work.  First, the
timescale for sedimentation and diffusion of heavy elements in the
accreted layer is often less than the accretion timescale
\citep{WWW82,BSW92,PBT07}.  \citet{PBT07} demonstrate that the
sedimentation of CNO ions can have a notable effect on
the ignition conditions of type I X-ray bursts at very low accretion
rates.  Second, our effective nuclear reaction network consists only
of hydrogen burning via the cold and hot CNO cycles and helium burning
via the triple-$\alpha$ process.  During a mixed hydrogen and helium
burst, however, hydrogen will burn predominantly via the $rp$-process of
\citet{WW81}.  Consequently, the light curves of these bursts would be
somewhat different than those shown in Figure \ref{Hint3.0} if 
$rp$-process hydrogen burning were included.  

A potentially serious issue that we are currently unable to address,
but that could have significant effects on the ignition of type I
X-ray bursts at low accretion rates, is the destruction of CNO ions
via nuclear spallation reactions \citep{BSW92}.  The accretion disks
around neutron stars that accrete at low rates are likely truncated
\citep[e.g.][]{D02,Ba04}.  Within the truncated disk, the accretion
flow is optically thin and quasi-spherical \citep[see, e.g.][]{NY95},
and thus the accreting ions have large radial velocities when they
reach the neutron star surface.  Observationally, this radiatively
inefficient accretion flow occurs in neutron star low-mass X-ray
binaries when $\dot{M} \lesssim 0.01 \dot{M}_{\mathrm{Edd}}$
\citep{Metal97,GD02}, the range of accretion rates applicable to this
work.  The directed kinetic energy of the accreting hydrogen and
helium may be sufficient to destroy nearly all of the ions in the
envelope that are heavier than helium.  The degree to which spallation
would affect type I X-ray burst ignition is unclear.  Compositional
inertia of the neutron star envelope could potentially counteract any
depletion of CNO due to spallation \citep{AJ82,WW84,TWWL93,Wetal04}.
However, if the heavy elements produced in a given burst sediment out
of the envelope before the next burst is triggered, compositional
inertia may be negligible, in which case spallation again becomes
important.  This issue in particular is worthy of further
investigation.

\acknowledgments

We would like to thank Edward Brown, Andrew Cumming, and Fang Peng for
useful discussions and the referee for several comments that helped us
improve this investigation.  This work was supported by NASA grant
NNG04GL38G.


\clearpage

\end{document}